\documentclass[11pt,a4paper,notitlepage]{article}
\usepackage[utf8]{inputenc}
\usepackage[english]{babel}
\usepackage[T1]{fontenc}
\usepackage{hyperref}
\usepackage{amsmath}
\usepackage{amsfonts}
\usepackage{color} 
\usepackage{caption}
\usepackage{array,multirow,makecell}
\setcellgapes{1pt}
\makegapedcells
\newcolumntype{R}[1]{>{\raggedleft\arraybackslash }b{#1}}
\newcolumntype{L}[1]{>{\raggedright\arraybackslash }b{#1}}
\newcolumntype{C}[1]{>{\centering\arraybackslash }b{#1}}
\newcommand{\Tr}{\mathrm{Tr}}

\newtheorem{proposition}{Proposition}

\newcommand{\sym}{\mathrm{Sym}}

\usepackage{enumerate}
\usepackage{subfig}
\usepackage{young}

\newcommand{\cG}{{\mathcal G}}

\newcommand{\cM}{{\mathcal M}}

\newcommand{\cT}{{\mathcal T}}

\newcommand{\cZ}{{\mathcal Z}}

\newcommand{\bP}{{\mathbf P}}
\newcommand{\bQ}{{\mathbf Q}}
\newcommand{\bS}{{\mathbf S}}

\definecolor{mygray}{gray}{0.3}

\newcommand\beq{\begin{equation}}
\newcommand\eeq{\end{equation}}
\newcommand{\bes}{\begin{eqnarray}}
\newcommand{\ees}{\end{eqnarray}}
\def\nn{{\nonumber}}

\def\vphi{{\varphi}}

\newcommand\restr[2]{{
  \left.\kern-\nulldelimiterspace 
  #1 
  \vphantom{\big|} 
  \right|_{#2} 
  }}
\def\extd{\mathrm {d}}

\usepackage{multicol}
\usepackage{amssymb}
\usepackage{graphicx}
\usepackage[left=2cm,right=2cm,top=2.5cm,bottom=2.5cm]{geometry}

\def\sym{\vcenter{\hbox{\begin{Young}  1  &  2 & 3 \cr \end{Young}}}}
\def\asym{\vcenter{\hbox{\begin{Young} 1  \cr  2 \cr 3 \cr \end{Young}}}}
\def\mixedb{\vcenter{\hbox{\begin{Young}  1  &  3 \cr 2 \cr \end{Young}}}}
\def\mixed{\vcenter{\hbox{\begin{Young}  1  &  2 \cr 3 \cr \end{Young}}}}

\begin{document}
\begin{center}
\textbf{\Large{{\centering Large $N$ limit of irreducible tensor models: \\
\medskip 
$O(N)$ rank-$3$ tensors with mixed permutation symmetry}
}}
\vspace{15pt}

{\large Sylvain Carrozza\footnote{\url{scarrozza@perimeterinstitute.ca}}}

\vspace{10pt}

{\sl Perimeter Institute for Theoretical Physics\\
 31 Caroline St N, Waterloo, ON N2L 2Y5, Canada\\
}

\end{center}

\vspace{5pt}

\begin{abstract}
\noindent It has recently been proven that in rank three tensor models, the anti-symmetric and symmetric traceless sectors both support a large $N$ expansion dominated by melon diagrams \cite{Benedetti:2017qxl}.
We show how to extend these results to the last irreducible $O(N)$ tensor representation available in this context, which carries a two-dimensional representation of the symmetric group $S_3$. 
Along the way, we emphasize the role of the irreducibility condition: it prevents the generation of vector modes which are not compatible with the large $N$ scaling of the tensor interaction. 
This example supports the conjecture that a melonic large $N$ limit should exist more generally for higher rank tensor models, provided that they are appropriately restricted to an irreducible subspace.   
\end{abstract}

\tableofcontents

\section{Introduction}

The first tensor models governed by a tractable $1/N$ expansion were discovered a few years ago \cite{expansion1, expansion2, expansion3}, and the methods developed in these early works have been generalized in various ways since then \cite{uncoloring, expansion4, expansioin5, expansioin6, Carrozza:2015adg}. Irrespectively of the rank $r\geq3$ of the tensor degrees of freedom, tensor models turn out to be generically dominated by a rather simple but non-trivial class of Feynman graphs, going under the name of \emph{melon diagrams} \cite{critical, review, RTM}. 

For some time, this new species of large $N$ expansion has mostly found applications as a tool to generate and understand random geometries in dimension $d > 2$, either in the context of tensor models themselves \cite{EDT, IsingD, melbp, Benedetti:2015ara, GurSch} or in group field theory \cite{BenGeloun:2011rc, Carrozza:2012uv, Samary:2012bw, tt2, Baratin:2013rja, Carrozza:2014rba, Delepouve:2015nia, Benedetti:2015yaa}. 

In the standard nomenclature of quantum field theories, the tensor models of random geometry are $0$-dimensional field theories. It is only recently that the large $N$ melonic limit of tensor degrees of freedom has started to be taken advantage of in the context of quantum mechanics and quantum field theory in more than one dimension. It has first been recognized by Witten \cite{Witten:2016iux} that the melonic large $N$ expansion of tensor models allows to emulate some properties of the celebrated SYK models \cite{Sachdev:1992fk, Kitaev, Maldacena:2016hyu, Polchinski:2016xgd} in the familiar context of large $N$ quantum mechanics (that is without disorder average). Another version of such models was then introduced by Klebanov and Tarnopolsky \cite{Klebanov:2016xxf}. Both models have been investigated in some detail since then \cite{Gurau:2016lzk, Krishnan:2016bvg, Krishnan:2017ztz, Krishnan:2017lra, Bonzom:2017pqs, Bulycheva:2017ilt, Benedetti:2018goh, Klebanov:2018nfp}, and have led to a variety of generalizations, including (but not restricted to): higher space-time dimensions \cite{Giombi:2017dtl, Prakash:2017hwq, Benedetti:2017fmp, BenGeloun:2017jbi}, higher order interactions \cite{Choudhury:2017tax}, and new asymptotic expansions for matrix-tensor models \cite{Ferrari:2017ryl, Azeyanagi:2017drg, Ferrari:2017jgw, Azeyanagi:2017mre}.

A key technical ingredient in all these instances of melonic behaviour is the presence of several independent copies of a symmetry group $G$ (e.g. $O(N)$ or $U(N)$), typically one for each index of the tensor, which greatly simplifies the combinatorial structure of the Feynman diagrams. In particular, it seemed crucial to impose no symmetry at all among the various indices of a given tensor. Because of its implications at the level of the Feynman amplitudes -- which can be indexed by colored cell complexes and colored graphs \cite{Ryan:2016sundry} -- the tensor models enjoying such a large symmetry have been going under the name of \emph{colored} or \emph{uncolored} models\footnote{Colored models, such as \cite{expansion1} and \cite{Witten:2016iux}, describe the dynamics of several species of fields, each labelled by a color. In contrast, uncolored models such as \cite{uncoloring} and \cite{Carrozza:2015adg, Klebanov:2016xxf} describe the dynamics of a single tensor, hence their name. However, both colored and uncolored models have a symmetry of the type $G^k$ ($k \geq 3$) resulting from the absence of symmetrization or antisymmetrization of their indices.}. 

The colored structure of the Feynman diagrams of colored and uncolored tensor models plays such a heavy role in the original proofs of the existence of their large $N$ expansion \cite{RTM}, that until recently it seemed unlikely that such results could be generalized to tensor models with less symmetry. However, motivated by the key conjecture and numerical evidence reported in \cite{Klebanov:2017nlk}, it was recently proven that both antisymmetric and symmetric traceless $O(N)$ rank-$3$ tensors support a melonic large $N$ expansion \cite{Benedetti:2017qxl}. The proof relies in part on methods first developed in \cite{Gurau:2017qya}, which successfully tackled a somewhat simpler tensor model involving two symmetric tensors. 

As one quickly realizes, a central technical ingredient of the arguments put forward in \cite{Klebanov:2017nlk, Benedetti:2017qxl} is the irreducibility of the subspace of tensors on which the model is based (indeed, antisymmetric and symmetric traceless tensors both carry an irreducible $O(N)$ representation). This observation suggests the following question: 
\begin{center}
\emph{Is the melonic large }$N$ \emph{limit generalizable to} irreducible tensors \emph{of arbitrary rank} $r \geq 3$\emph{?}
\end{center}
In this paper, we make a step in the direction of an affirmative answer. We focus on the third available class of rank-$3$ irreducible tensor spaces, whose elements transform as a two-dimensional representation of the permutation group $S_3$. We show that the methods and results of \cite{Benedetti:2017qxl} do generalize to this last rank-$3$ invariant space, providing another instance of melonic large $N$ behaviour in tensor models and tensor field theories. 

\medskip

The paper is organized as follows. In section \ref{sec:model}, we first introduce the class of rank-$3$ tensor models we wish to consider, and identify a potential instability triggered by the trace modes of the tensor. We then define more precisely the mixed symmetry model we will consider in the later sections, in which such trace modes have been correctly removed. In section \ref{sec:feynman}, we introduce a perturbative expansion of the model, indexed by Feynman maps and stranded graphs. Following \cite{Benedetti:2017qxl}, in section \ref{sec:tadmel} we identify and proceed to a resummation of the infinite family of melon-tadpole maps. This allows to introduce a new perturbative expansion in section \ref{sec:largeN}, which is proven to admit a well-defined $1/N$ expansion dominated by melon maps. We close with a conclusion and some general comments. 

\section{Irreducible O(N) tensor models in rank three}\label{sec:model}

We introduce $O(N)$ rank three tensor models in general, and define more precisely the theory with mixed symmetry which is the main subject of this note.  

\subsection{Rank-3 tensor models with complete interaction}

We consider a bosonic theory in zero dimension with $O(N)$ global symmetry. The degrees of freedom are organized into a real rank-$3$ tensor $T_{a_1 a_2 a_3} $ transforming as a product of three fundamental representations\footnote{We assume Einstein's summation convention throughout the paper.}:
\beq\label{eq:On-action}
\forall O \in O(N)\,, \quad  T_{a_1 a_2 a_3} \rightarrow [O \cdot T]_{a_1 a_2 a_3} := O_{a_1 b_1} O_{a_2 b_2} O_{a_3 b_3} T_{b_1 b_2 b_3}  \,. 
\eeq
The scalar product $\langle \cdot \vert \cdot \rangle$ is the standard one, namely: $\langle T \vert T'\rangle = T_{a_1 a_2 a_3} T'_{a_1 a_2 a_3}$. 

The partition function is taken to be of the general form\footnote{We adopt the standard sign convention of matrix models. Note that the interaction as no reason to be bounded. At this stage, $\cZ_{N} (\lambda)$ is considered as a formal power series in $\lambda$. But as usual, the leading order sector at large $N$ will eventually turn out to be summable. }:
\beq\label{eq:partition-function}
\cZ_{N} (\lambda) := \int \extd \mu_\bP (T) \, \exp\left( \frac{\lambda}{4 N^{3/2}} T_{i_1 i_2 i_3} T_{i_3 i_4 i_5} T_{i_5 i_2 i_6} T_{i_6 i_4 i_1} \right) \,.
\eeq
where $\bP$ is an $O(N)$-invariant covariance which will be fixed in subsection \ref{sec:mixed-model}. The $N^{-3/2}$ scaling of the interaction is the standard one, and is necessary to the existence of a melonic large $N$ expansion with 't Hooft coupling $\lambda$. Remark however that the quartic interaction -- which is invariant under $O(N)$, as it should -- is not invariant under a larger $O(N)^3$ symmetry, so that our Ansatz can not fall in the class of colored tensor models studied in \cite{Carrozza:2015adg}.

\begin{figure}[htb]
 \begin{center}
\includegraphics[scale=.7]{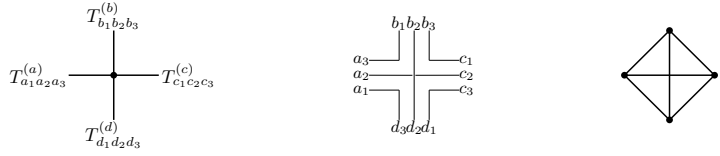}
 \caption{Graphical representations of the interaction kernel: as an ordinary (embedded) vertex (left); and as a stranded diagram (middle). The associated boundary graph (right) is the complete graph on four vertices.} \label{fig:vertex}
 \end{center}
 \end{figure}

The quartic interaction defines a map:
\beq
V(T^{(a)}, T^{(b)}, T^{(c)}, T^{(d)}) 
= V_{a_1 a_2 a_3 , b_1 b_2 b_3, c_1 c_2 c_3, d_1 d_2 d_3}  T^{(a)}_{a_1 a_2 a_3} T^{(b)}_{b_1 b_2 b_3} T^{(c)}_{c_1 c_2 c_3} T^{(d)}_{d_1 d_2 d_3}\,,
\eeq
with kernel
\beq\label{kernel-int}
V_{a_1 a_2 a_3 , b_1 b_2 b_3, c_1 c_2 c_3, d_1 d_2 d_3} := \delta_{a_3 b_1} \delta_{b_3 c_1} \delta_{c_3 d_1} \delta_{d_3 a_1} \delta_{a_2 c_2} \delta_{b_2 d_2}
\eeq
As standard in the literature, we will represent the interaction kernel as an ordinary four-valent vertex, or as a stranded diagram -- as shown in Fig. \ref{fig:vertex}. In the stranded representation, each strand represents a contraction by a Kronecker delta, according to \eqref{kernel-int}. As a result of this non-trivial combinatorial structure, the representation of the vertex kernel as an ordinary vertex is sensitive to its embedding in the plane. One can also represent such an interaction by the boundary graph\footnote{See e.g. \cite{Benedetti:2017qxl} for a general definition of boundary graph in tensor models. We will not make further use of this notion in the present paper.} shown on the right panel of Fig. \ref{fig:vertex}. As this is the complete graph on four vertices, we propose to call such an interaction a \emph{complete interaction}\footnote{It is also sometimes known as a tetrahedral interaction, since the complete graph on four vertices is the one-skeleton of a tetrahedron.}. By this definition, there exists other complete interactions (for instance the $O(N)^3$-invariant interaction of \cite{Carrozza:2015adg}) that one might want to include in the action. However, they all become equivalent upon reduction to a tensor in a fixed irreducible representation of $S_3$, so it is sufficient for our purpose to consider only one such interaction. 

\subsection{Vector modes and trace instability}\label{sec:vectormodes}

As we now explain, not all choices of propagator $\bP$ (with reasonable scaling in $N$) lead to an interesting large $N$ expansion. For instance, let us consider the natural (but eventually unfortunate) choice of free propagator:
\beq\label{eq:naive-propa}
\langle T_{a_1 a_2 a_3} T_{b_1 b_2 b_3} \rangle_{0} = \int \extd \mu_{\bP} (T) \, T_{a_1 a_2 a_3} T_{b_1 b_2 b_3} = \delta_{a_1 b_1} \delta_{a_2 b_2} \delta_{a_3 b_3}\,,
\eeq
corresponding to a free action $S_{\mathrm{free}} = \frac{1}{2} \langle T \vert T \rangle$. This theory leads to amplitudes which diverge arbitrarily fast in $N$ and therefore can not support a well-behaved melonic large $N$ expansion. As we will see shortly, the problem can be boiled down to the presence of vector modes arising from the traces of the tensor $T$, which couple in an ill-behaved way to the traceless part of $T$.  

In more detail, one can decompose the tensor $T_{a_1 a_2 a_3}$ into:
\beq
T_{a_1 a_2 a_3} = T^0_{a_1 a_2 a_3} + \frac{1}{\sqrt{N}} \left( \chi_{a_1} \delta_{a_2 a_3} + \vphi_{a_2} \delta_{a_1 a_3} + \psi_{a_3} \delta_{a_1 a_2} \right) \,,
\eeq 
where $T^0$ is a completely traceless tensor ($T^0_{aab} = T^0_{aba} = T^0_{baa}$), while $\chi$, $\vphi$ and $\psi$ are vectors which can be expressed in terms of the three traces of $T$. For instance, one finds:
\beq
\vphi_i = \frac{\sqrt{N}}{N^2 + N - 2} \left(- T_{aai} + (N+1) T_{aia} - T_{iaa} \right)\,.
\eeq
With the naive covariance \eqref{eq:naive-propa}, the propagator of $\vphi$ at leading order in $1/N$ is appropriately normalized (up to $1/N$ corrections) 
\beq
\langle \vphi_i \vphi_j \rangle_0 \sim \delta_{ij}\,.
\eeq 

The vector modes interact among themselves, and also with $T^0$. We provide two examples, one of which does spoil the large $N$ structure of the model. 

\begin{figure}[htb]
 \begin{center}
\includegraphics[scale=.7]{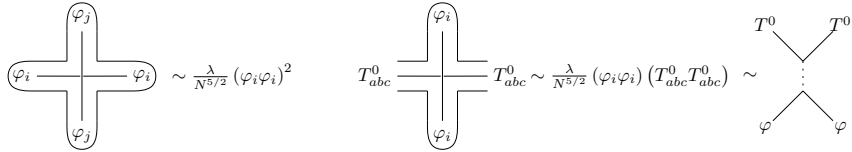}
 \caption{Graphical representations of two of the interactions involving the vector $\vphi_i$: the self-interaction on the left is sufficiently suppressed in $1/N$, while the double-trace interaction on the right leads to an instability.} \label{fig:int_vectors}
 \end{center}
 \end{figure}
 
Let us start with the induced $\vphi^4$ self-interaction:
\beq
\frac{\lambda}{N^{3/2}} \left( \frac{1}{\sqrt{N}} \right)^4 N \left( \vphi_i \vphi_i \right)^2 \sim \frac{\lambda}{N^{5/2}} \left( \vphi_i \vphi_i \right)^2 \,,
\eeq
with one factor of $N$ coming from the contraction of delta functions represented on the left panel of Fig. \ref{fig:int_vectors}. 
Since the scaling leading to a sensible large $N$ expansion for such a vector interaction is $1/N$ \cite{Moshe:2003xn}, this particular coupling does not preclude the existence of a melonic large $N$ expansion. 

\begin{figure}[htb]
 \begin{center}
\includegraphics[scale=.7]{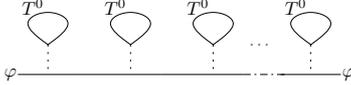}
 \caption{Chains of tadpole loops leading to $\vphi_i$ two-point functions of arbitrary high order in $N$.} \label{fig:tadpoles_vectors}
 \end{center}
 \end{figure}

However, there are also more problematic couplings between $T^0$ and the vector degrees of freedom, such as:
\beq
\frac{\lambda}{N^{3/2}} \left( \frac{1}{\sqrt{N}} \right)^2 \vphi_i \vphi_i  T_{abc}^0 T_{abc}^0 \sim \frac{\lambda}{N^{5/2}}  \left(\vphi_i \vphi_i \right)  \left( T_{abc}^0 T_{abc}^0 \right) \,.
\eeq
See again Fig. \ref{fig:int_vectors} for a graphical representation.
This double-trace interaction does generate arbitrary large powers of $N$, for instance through chains of tadpole graphs, as depicted in Fig. \ref{fig:tadpoles_vectors}. Each tadpole loop in such a diagram generates a factor of $N^3$ which is not entirely balanced by the $1/N^{5/2}$ scaling of the interaction. This leads to amplitudes scaling as $\sim N^{p/2}$, where $p$ is the number of tadpoles in the chain. In \cite{Benedetti:2017qxl}, this type of pathological behaviour is referred to as a \emph{trace instability}, and has been explicitly identified in a symmetric (but not traceless) tensor model. 

There are two ways of curing such problems. One is to introduce as many independent new \hbox{'t Hooft} couplings as needed to tame the problematic large $N$ contributions\footnote{I would like to thank E. Witten for pointing this out.}. For instance, one would need to include a new independent coupling to parametrize the "renormalized" interaction $\left(\vphi_i \vphi_i \right)  \left( T_{abc}^0 T_{abc}^0 \right)$. This procedure might lead to the definition of interesting vector-tensor models. For the purpose of the present paper, we prefer to stick to pure tensor models. We therefore adopt a simpler strategy, consisting in choosing $\bP$ such that no vector modes at all can propagate in the model. Imposing the traceless condition by itself is not necessarily sufficient (since mixing between sub-representations may possibly generate such vector contributions), but working with an irreducible representation is. 

\subsection{Irreducible tensors}

For the convenience of the reader, we briefly review how the tensor product of three fundamental $O(N)$ representations decomposes into irreducible representations (see e.g. \cite{weyl1946, hamermesh} for additional details). There are two operations which obviously commute with the $O(N)$ action \eqref{eq:On-action}: 1) taking the trace over two indices; and 2) permuting the indices in an arbitrary way. Moreover, it turns out they are the only independent transformations which needs to be reduced.

Let us first discuss the action of the permutation group $S_3$, which we take to be:
\beq
\forall \sigma \in S_3 \,, \qquad \left(\sigma \triangleright T \right)_{a_1 a_2 a_3} := T_{a_{\sigma(1)} a_{\sigma(2)} a_{\sigma(3)}}\,.  
\eeq
The irreducible representations of $S_3$ are labelled by Young tableaux with three boxes. One therefore associates a Young tableau to each tensor subspace with fully reduced permutation symmetry.  Graphically, one has:
\beq\label{eq:young}
\vcenter{\hbox{\begin{Young}  1 \cr \end{Young}}} \otimes \mkern-18mu \vcenter{\hbox{\begin{Young}  2 \cr \end{Young}}} \otimes \mkern-18mu \vcenter{\hbox{\begin{Young}  3 \cr \end{Young}}}  \quad = \sym \quad \oplus \asym \quad \oplus \mixedb \quad \oplus \mixed
\eeq
In this representation, a box \kern-1em $\vcenter{\hbox{\begin{Young}  i \cr \end{Young}}}$ stands for the i$^{\mathrm{th}}$ index of the tensor. A Young operator is then uniquely associated to each Young tableau generated on the right-hand side of the equality: it imposes a symmetrization over the indices appearing in a same row, followed by an antisymmetrization of the indices appearing in a same column. In particular, the Young tableau with a single row (resp. column) corresponds to the subspace of completely symmetric (resp. antisymmetric) tensors. The last two sub-representations have mixed symmetry, in the sense that they carry a two-dimensional representation of $S_3$. They are furthermore equivalent. Finally, the dimensions of these vector spaces are:
\begin{align}
\mathrm{dim} \left( \; \mkern-18mu \sym \; \right) &= \frac{1}{6} N \left( N^2 + 3 N + 2 \right) \,, \nn \\
\mathrm{dim} \left( \; \mkern-18mu \asym \; \right)  &= \frac{1}{6} N \left( N-1 \right) \left( N- 2 \right) \,, \\
\mathrm{dim} \left( \; \mkern-18mu \mixed \; \right) &= \mathrm{dim} \left(\; \mkern-18mu \mixedb \; \right) = \frac{1}{3} N \left( N^2 - 1 \right) \,. \nn
\end{align}
In the rest of the paper, we will construct a tensor model with mixed symmetry. For definiteness, we will focus on the tableau $\mkern-18mu \mixed$. Its associated Young operator $\bS$ acts on an arbitrary tensor $T$ as: 
\beq
(\bS T)_{a_1 a_2 a_3} := \frac{1}{3} \left( \left(1 - (1\, 3) \right) \left(1 + (1\, 2) \right) \triangleright T \right)_{a_1 a_2 a_3} = \frac{1}{3} \left( T_{a_1 a_2 a_3} + T_{a_2 a_1 a_3} - T_{a_3 a_2 a_1} - T_{a_2 a_3 a_1} 
\right) \,,
\eeq
yielding a tensor which is in particular anti-symmetric in $a_1 \leftrightarrow a_3$.
The kernel of $\bS$ is:
\begin{align}
\bS_{a_1 a_2 a_3, b_1 b_2 b_3} := \frac{1}{3} \left( \delta_{a_1 b_1} \delta_{a_2 b_2} \delta_{a_3 b_3} + \delta_{a_1 b_2} \delta_{a_2 b_1} \delta_{a_3 b_3} - \delta_{a_1 b_3} \delta_{a_2 b_2} \delta_{a_3 b_1} - \delta_{a_1 b_3} \delta_{a_2 b_1} \delta_{a_3 b_2} 
\right) \,.
\end{align}

In order to reduce to an irreducible tensor representation with order $N^3$ degrees of freedom, one must further remove the trace modes (which contain only $N$ degrees of freedom each). There are three of them: one lies in the completely symmetric part, the other two are in the mixed sectors. For these three sectors, the reduction is completed by acting with the orthogonal projector on traceless tensors:
\begin{align}
\bQ_{a_1 a_2 a_3, b_1 b_2 b_3} & := \delta_{a_1 b_1} \delta_{a_2 b_2} \delta_{a_3 b_3} \nn \\
&- \frac{\delta_{a_1 a_2}}{N^2 + N - 2} \left( \left( N+1 \right) \delta_{a_3 b_3} \delta_{b_1 b_2} - \delta_{a_3 b_2} \delta_{b_1 b_3} - \delta_{a_3 b_1} \delta_{b_2 b_3}  \right) \nn \\
&- \frac{\delta_{a_1 a_3}}{N^2 + N - 2} \left( - \delta_{a_2 b_3} \delta_{b_1 b_2} +  \left( N+1 \right) \delta_{a_2 b_2} \delta_{b_1 b_3} - \delta_{a_2 b_1} \delta_{b_2 b_3}  \right)  \nn \\
&- \frac{\delta_{a_2 a_3}}{N^2 + N - 2} \left( - \delta_{a_1 b_3} \delta_{b_1 b_2} - \delta_{a_1 b_2} \delta_{b_1 b_3} + \left( N+1 \right)  \delta_{a_1 b_1} \delta_{b_2 b_3}  \right) 
\end{align}

Hence, we have exactly three irreducible tensor representations at our disposal. The symmetric traceless and antisymmetric sectors have already been treated elsewhere \cite{Benedetti:2017qxl}. The irreducible representation associated to the tableau $\mkern-18mu \mixed$ (which is equivalent to that of the tableau $\mkern-18mu \mixedb$) is studied in the remainder of the present paper.

\subsection{Tensor model with mixed permutation symmetry}\label{sec:mixed-model}

In the decomposition \eqref{eq:young}, the two representations with mixed symmetry are not orthogonal to each other with respect to $\langle \cdot \vert \cdot \rangle$\footnote{This is in contrast with the symmetric and antisymmetric sectors, which are decomposed in an orthogonal manner.}. This implies that the projector $\bS$ is not symmetric, and therefore cannot immediately be used as a propagator. 
One may instead consider
\beq
\tilde{\bP} := \frac{3}{4}  \bS \bS^\top\,. 
\eeq
The $3/4$ coefficient has been chosen to ensure that, not only $\tilde{\bP}^\top = \tilde{\bP}$, but also $\tilde{\bP}^2 = \tilde{\bP}$. Hence $\tilde{\bP}$ is an orthogonal projector whose image is included in the image of $\bS$. One may furthermore check that 
\beq
\Tr \, \tilde{\bP} = \frac{1}{3} N \left( N^2 - 1 \right) = \mathrm{dim} \left( \; \mkern-18mu \mixed \; \right) = \mathrm{rk} \, \bS \,.
\eeq
Hence $\tilde{\bP}$ is nothing but the orthogonal projector on the subspace of tensors associated to the tableau $\mkern-18mu \mixed$. More explicitly, we have:
\begin{align}
\tilde{\bP}_{a_1 a_2 a_3, b_1 b_2 b_3} &= \frac{1}{3} \left( \delta_{a_1 b_1} \delta_{a_2 b_2} \delta_{a_3 b_3}  - \delta_{a_1 b_3} \delta_{a_2 b_2} \delta_{a_3 b_1} \right) \nn \\
& + \frac{1}{6} \left( \delta_{a_1 b_2} \delta_{a_2 b_1} \delta_{a_3 b_3}  + \delta_{a_1 b_1} \delta_{a_2 b_3} \delta_{a_3 b_2} \right) \\
& -  \frac{1}{6} \left( \delta_{a_1 b_2} \delta_{a_2 b_3} \delta_{a_3 b_1}  + \delta_{a_1 b_3} \delta_{a_2 b_1} \delta_{a_3 b_2} \right) \nn
\end{align}

We finally define the propagator of the model as:
\beq
\bP := \bQ \tilde{\bP} \bQ = \tilde{\bP} \bQ  = \bQ \tilde{\bP}\,,
\eeq
which is the orthogonal projector on traceless tensors with symmetry $\mkern-18mu \mixed$. A direct calculation shows that:
\begin{align}\label{eq:propa}
\bP_{a_1 a_2 a_3, b_1 b_2 b_3} &= \frac{1}{3} \left( \delta_{a_1 b_1} \delta_{a_2 b_2} \delta_{a_3 b_3}  - \delta_{a_1 b_3} \delta_{a_2 b_2} \delta_{a_3 b_1} \right) \nn \\
& + \frac{1}{6} \left( \delta_{a_1 b_2} \delta_{a_2 b_1} \delta_{a_3 b_3}  + \delta_{a_1 b_1} \delta_{a_2 b_3} \delta_{a_3 b_2} \right) \nn \\
& -  \frac{1}{6} \left( \delta_{a_1 b_2} \delta_{a_2 b_3} \delta_{a_3 b_1}  + \delta_{a_1 b_3} \delta_{a_2 b_1} \delta_{a_3 b_2} \right)  \\
&+ \frac{1}{2(N-1)} \left( \delta_{a_1 b_3} \delta_{a_2 a_3} \delta_{b_1 b_2} 
+  \delta_{a_1 a_2} \delta_{a_3 b_1} \delta_{b_2 b_3} \right) \nn \\
&- \frac{1}{2(N-1)} \left( \delta_{a_1 b_1} \delta_{a_2 a_3} \delta_{b_2 b_3} +  \delta_{a_1 a_2} \delta_{a_3 b_3} \delta_{b_1 b_2} \right) \nn
\end{align}
The Gaussian measure $\extd \mu_\bP$ is degenerate but defines a suitable covariance:
\beq
\int \extd \mu_\bP (T) \, T_{a_1 a_2 a_3} T_{b_1 b_2 b_3} = \bP_{a_1 a_2 a_3, b_1 b_2 b_3} \,.
\eeq

\begin{figure}[htb]
 \begin{center}
\includegraphics[scale=.7]{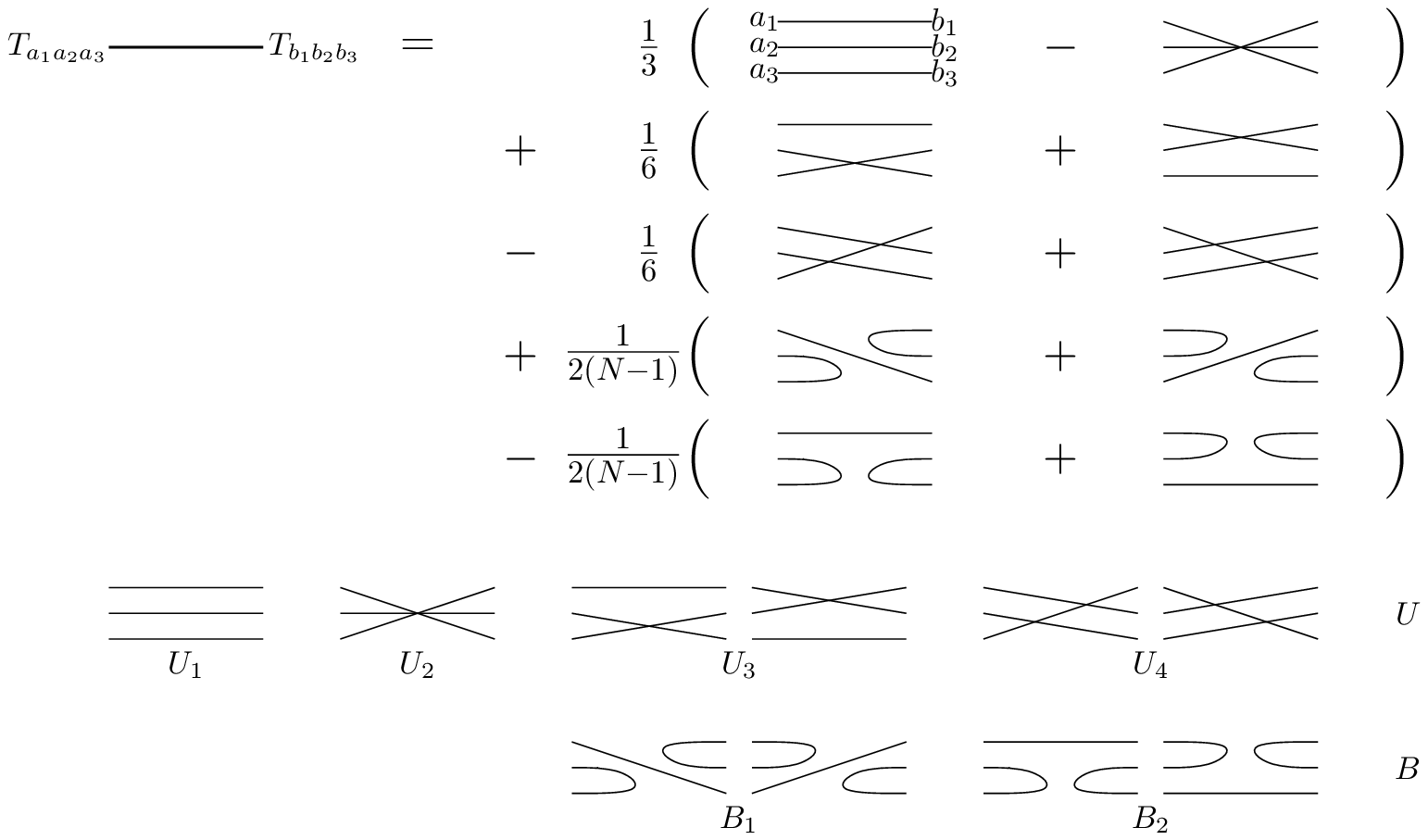}
 \caption{Graphical representation of the propagator \eqref{eq:propa}. Stranded configurations have been organized into several classes of broken ($B$) and unbroken ($U$) configurations.} \label{fig:propa}
 \end{center}
 \end{figure}

As shown in Fig. \ref{fig:propa}, it is convenient to represent the propagator as a plain line, which decomposes further into triplets of strands. Each triplet represents one of the ten patterns of Kronecker delta contractions appearing in \eqref{eq:propa}, analogously to the representation adopted for the vertex.
We will call \emph{broken} (resp. \emph{unbroken}) the last four (resp. the first six) terms of the propagator, and will label them accordingly by the letter $B$ (resp. $U$). More precisely, we will separate the different broken and unbroken configurations into subclasses $B_1, \, B_2, \, U_1 , \, \ldots,\, U_4$, as shown in Fig. \ref{fig:propa}.  

\section{Feynman expansion and amplitudes}\label{sec:feynman}

The Feynman expansion is indexed by combinatorial objects associated to gluings of propagator lines and interaction vertices. We distinguish between the plain line representation, which gives rise to \emph{Feynman maps}, and the more fine-grained stranded representation, which yields \emph{stranded graphs}.

\subsection{Feynman maps}

The Feynman maps are the combinatorial objects which directly label the various terms obtained by Wick contraction in the perturbative Feynman expansion.  
Since the interaction vertex is naturally equipped with a cyclic order -- or equivalently a local embedding into the plane --, the theory is most naturally expanded in terms of \emph{combinatorial maps}\footnote{A combinatorial map is a formalization of the notion of graph embedded into an orientable surface -- or equivalently a ribbon graph. The reader is referred to e.g. \cite{lando-zvonkin-graphs, ellis2013graphs} for further details about these standard definitions.}. In the present model the combinatorial maps of interest are furthermore $4$-regular, a qualifier we will keep implicit in the rest of the paper.

In particular, the full connected two-point function can be expanded as a sum over amplitudes of connected two-point maps: 
\beq\label{connected2point}
C_{a_1 a_2 a_3 , b_1 b_2 b_3} := \langle T_{a_1 a_2 a_3} T_{b_1 b_2 b_3}\rangle_{\mathrm{c}} = \sum_{\mathrm{connected}\; 2\mathrm{-point} \; \mathrm{maps}\;\cM} \lambda^{V(\cM)} A(\cM)_{a_1 a_2 a_3 , b_1 b_2 b_3} \,.
\eeq
Its trace $C:= C_{a_1 a_2 a_3 , a_1 a_2 a_3}$ in turn expands as a sum over vacuum \emph{rooted} maps\footnote{A rooted map is a map with one edge marked by an arrow.}:
\beq\label{trace2point}
C = \frac{N \left( N^2 - 4 \right)}{3} + \sum_{\mathrm{connected}\; \mathrm{rooted} \; \mathrm{maps}\;\cM} \lambda^{V(\cM)} A(\cM)\,,
\eeq
where the first term is the trace of the free propagator $\bP$\footnote{In the language of \cite{Benedetti:2017qxl}, it is the contribution of the \emph{ring diagram}.}.

We denote by $V(\cM)$ (resp. $E(\cM)$) the number of vertices (resp. edges) in a Feynman map $\cM$. If $\cM$ is a vacuum map, its $4$-regular nature implies that $2 V (\cG) = E (\cG)$.

Finally, we note that the vertex is invariant under any permutation that preserves the cyclic order of its external legs up to orientation. For instance:
\beq
V(T^{(a)}, T^{(b)}, T^{(c)}, T^{(d)}) = V( T^{(b)}, T^{(c)}, T^{(d)}, T^{(a)} ) = V( T^{(a)}, T^{(d)}, T^{(c)}, T^{(b)} )\,,
\eeq   
which may be represented pictorially as:
\begin{center}
\includegraphics[scale=.7]{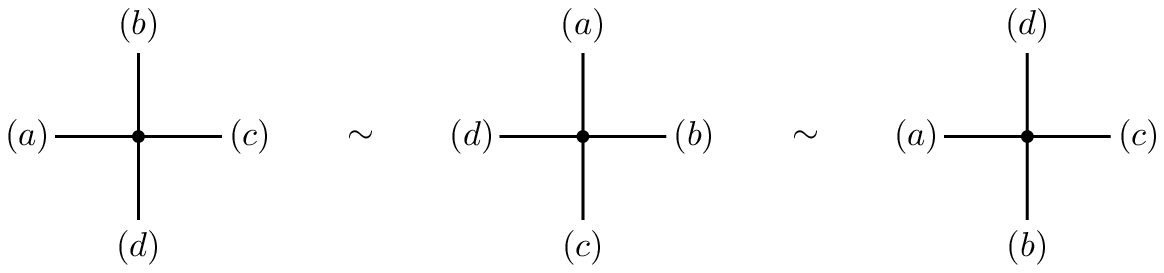}
\end{center}
We also remark that, in contrast to the antisymmetric and symmetric traceless models, the vertex is not invariant under the full permutation group of its legs. Hence, it is essential to keep track of the embedding information encoded in a Feynman map. 
This being said, we will often implicitly identify two Feynman maps which are in the same equivalent class under reversal of the cyclic orders around their vertices\footnote{Such intermediate combinatorial objects, which are neither abstract graphs nor maps, are sometimes referred to as \emph{cyclically ordered graphs} \cite{ellis2013graphs}. While maps can be thought of as ribbon graphs, which are collections of discs connected by ribbons, cyclically ordered graphs can be thought of as collections of discs connected by strings.}, since these have the same amplitude.

\subsection{Stranded graphs}

Each vacuum map $\cM$ further decomposes into $10^{2V(\cM)}$ \emph{stranded graph}\footnote{Note that the stranded graphs we discuss in this paper are always \emph{vacuum} graphs.} configurations $G$, each corresponding to a choice of one among ten terms of \eqref{eq:propa} for every edge of $\cM$. The patterns of identifications associated to a choice of stranded configuration results in a collection of closed cycles of strands, which are called \emph{faces}. We denote by $\hat{G}(\cM)$ the set of stranded configurations of $\cM$, and by $F(G)$ (resp. $B(G)$, $U(G)$, etc.) the number of faces (resp. broken edges, unbroken edges, etc.) of $G \in \hat{G}(\cM)$. The stranded graphs come with a definite scaling in $N$, each face bringing one factor of $N$. More precisely, it is easy to show that:
\beq
A(\cM) = \sum_{G \in \hat{G}(\cM)} \frac{\epsilon(G)}{R(G)} N^{3- \omega(G)}\,,
\eeq  
where $\omega(G)$ is the \emph{degree} of the stranded configuration $G$
\beq
\omega(G) = 3 + \frac{3}{2} V(G) + B(G) - F(G)\,,
\eeq
while $\epsilon(G)$ and $R(G)$ are respectively a sign and a normalization factor of order $N^0$
\begin{align}
\epsilon &= (-1)^{U_2  + U_4  + B_2 } \,,  \\
R &= 3^{U_1 + U_2} 6^{U_3+U_4} \left( 2(1-1/N) \right)^{B}\,. \nn
\end{align}

\begin{figure}[htb]
 \begin{center}
\includegraphics[scale=.7]{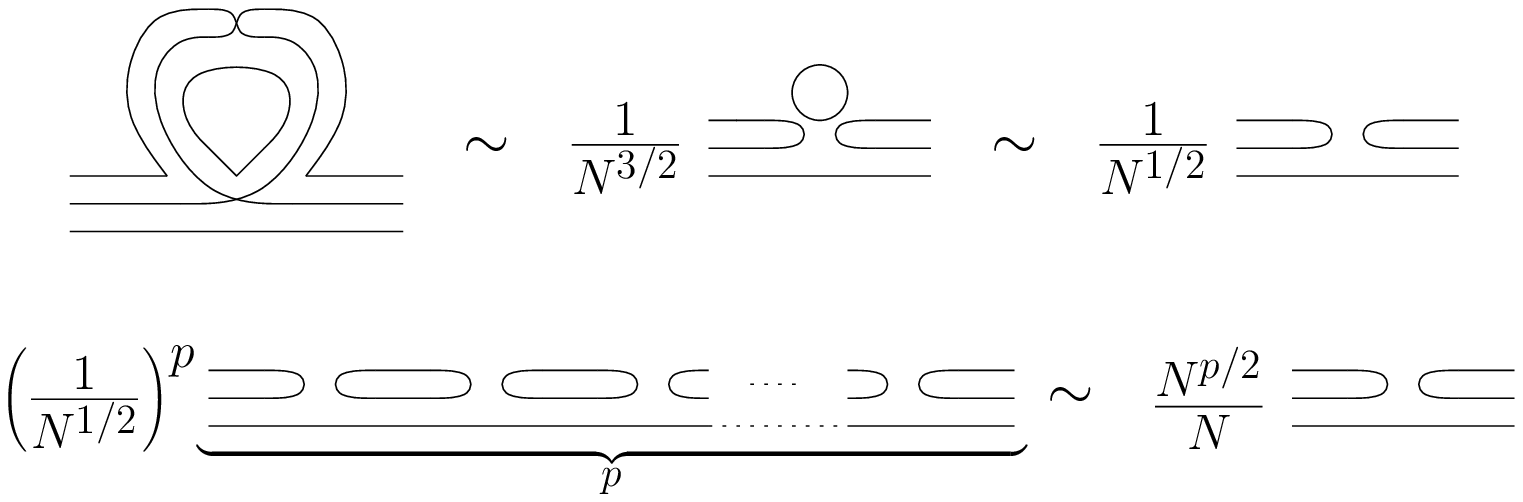}
 \caption{Chains of bad tadpole configurations, yielding unbounded degrees.} \label{fig:bad-tadpoles}
 \end{center}
 \end{figure}

The main subtlety in such models -- in comparison with their simpler colored and uncolored cousins  -- is that not all stranded graphs have positive degree. For instance, one may generate arbitrarily negative degrees by chaining tadpoles as represented in Fig. \ref{fig:bad-tadpoles}. The specific tadpole configurations giving rise to such pathological amplitudes have earned the name of \emph{bad tadpoles} in \cite{Benedetti:2017qxl}.  
 
Hence the existence of the $1/N$ expansion must rely on delicate cancellations. Given our general discussion of section \ref{sec:vectormodes}, where we emphasized the importance of removing the trace modes, this feature is of course to be expected. A quick inspection of Fig.~\ref{fig:bad-tadpoles} shows that the chains of bad tapoles do indeed correspond to the propagation of trace modes, and are essentially identical to those of Fig.~\ref{fig:tadpoles_vectors}. We therefore expect them to be correctly compensated by terms with opposite signs in the present traceless model, which we will confirm explicitly.  

In more detail, we follow the general method introduced in \cite{Benedetti:2017qxl}, and introduce a family of maps requiring special attention: the melon-tadpole maps. An \emph{elementary melon} is defined as a connected two-point map with two vertices, while a \emph{tadpole} is simply a two-point map with a single vertex. We then call \emph{melon} (or \emph{melonic}) any connected two-point map obtained from the bare propagator by recursively replacing edges of the graph by elementary melons. We similarly call \emph{melon-tadpole} any connected two-point map obtained by recursively replacing edges by elementary melons or tadpoles. Finally, we will also call \emph{generalized tadpole} (resp. \emph{generalized melon}) a tadpole (resp. melon) in which one of the internal edges has been replaced by a non-trivial two-point graph. These notions are important  because of the following proposition, which we recall without proof. 

\begin{proposition}[\cite{Benedetti:2017qxl}, propositions 3 and 4]\label{propo1}
Let $G$ be a connected stranded graph. 
\begin{enumerate}[i)]
\item If $G$ has no melon and no tadpole, then $\omega(G) \geq 0$.
\item If $G$ has no generalized melon and no generalized tadpole, then $\omega(G) \geq 1/2$.
\end{enumerate}
\end{proposition}

To prove that the model defined in \eqref{eq:partition-function} and \eqref{eq:propa} admits a large $N$ expansion dominated by melon maps, we can follow \cite{Benedetti:2017qxl} and proceed in three steps:
\begin{enumerate}
\item Prove that the family of melon-tadpoles can be resummed in a controlled manner, and that its sum reduces to the melonic two-point function in the large $N$ limit.  
\item Define a new perturbative expansion indexed by maps with no melons and no tadpoles, which by Proposition \ref{propo1} \emph{i)} will admit a well-defined large $N$ expansion.
\item Prove that generalized tadpoles cannot be leading order; use Proposition \ref{propo1} \emph{ii)} and item 1 to deduce that the full two-point function of the model reduces to the melonic two-point function in the large $N$ limit. 
\end{enumerate}

\section{Resummation of the melon-tadpole family}\label{sec:tadmel}

In this section we resum the infinite family of melon-tadpole two-point maps, and show that the resulting two-point function is a simple renormalized version of the bare propagator. 

\subsection{General form of a two-point function}

To begin with, we remark that because $\bP$ is an orthogonal projector onto an irreducible $O(N)$ tensor representation, any two-point map must have an amplitude proportional to the bare propagator. 
Indeed, by construction any two-point map defines an $O(N)$ intertwiner which leaves the space of irreducible $\mkern-18mu \, \mixed$ tensors invariant. By Schur's Lemma it is therefore proportional to the identity on this subspace. Since it is guaranteed to vanish on the orthogonal complement, it must thus be proportional to the projector $\bP$ itself.

Graphically, we simply have:
\begin{center}
\includegraphics[scale=.8]{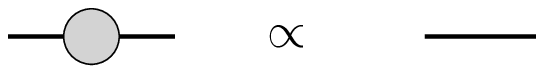}
\end{center}

Let us denote by $\mathbf{K}$ the melon-tadpole two-point function. Our simple observation guarantees the existence of a function $K(\lambda , N)$ such that:
\beq
\mathbf{K} = K(\lambda , N) \, \bP\,.
\eeq
In the remainder of this section, we will prove that $K(\lambda, N)$ is a series in $\frac{1}{\sqrt{N}}$ and therefore admits a well-defined large $N$ expansion. To get there, we first need to compute the amplitudes of tadpoles and elementary melons. 

\subsection{Tadpole maps}

We start by analyzing the tadpole contributions. There are two inequivalent tadpole Feynman maps:
\begin{center}
\includegraphics[scale=.7]{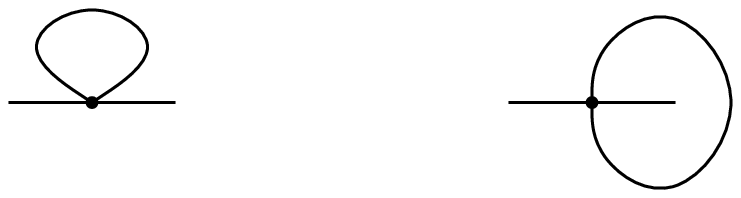}
\end{center}
The first (which we call planar) has multiplicity $2$, the second (which we call non-planar) has multiplicity $1$. 

The amplitude of a planar tadpole can be shown to evaluate to
\beq
A( \vcenter{\hbox{\includegraphics[scale=0.4]{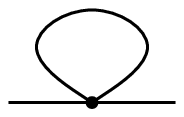}}} ) = \lambda \frac{ (N - 2)(4N-1)}{12 N^{3/2} (N-1)} \bP_{a_1 a_2 a_3 , b_1 b_2 b_3}\,,
\eeq
while that of a non-planar tadpole yields
\beq
A(\vcenter{\hbox{\includegraphics[scale=0.4]{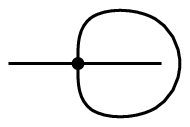}}}) = \lambda \frac{N^2 - 4}{3 N^{3/2} (N-1)} \bP_{a_1 a_2 a_3 , b_1 b_2 b_3}\,.
\eeq
We remark that both amplitudes have prefactors scaling as $\frac{1}{N^{1/2}}$, as is to be expected: indeed, given that such maps have only one cycle, their stranded configurations can have at most one face, yielding a maximal scaling $\frac{N}{N^{3/2}}$. More interestingly, the broken configurations of $\bP$ are weighted by an overall factor $\sim \frac{1}{N^{3/2}}$, which is to be contrasted with the $\frac{1}{N^{1/2}}$ we would naively expect from bad tadpole configurations. These problematic configurations simply average to zero, as we were hoping for. Having recognized that $A(\cG)$ must be proportional to $\bP$, we could actually have anticipated it, without explicitly computing the amplitudes. The scaling of broken edges in $\bP$ being always suppressed by a factor $\sim \frac{1}{N}$ with respect to the unbroken ones, it suffices to realize that the unbroken strands are at most of order $N^{-1/2}$ to conclude that the broken ones must combine in such a way that they do not contribute before order $N^{-3/2}$. 

We finally introduce a function $f_T (N)$ parametrizing the total contribution of tadpole maps: 
\beq
2 A ( \vcenter{\hbox{\includegraphics[scale=0.4]{planar-tadpole}}} ) + A (\vcenter{\hbox{\includegraphics[scale=0.4]{non-planar-tadpole}}}) = \lambda \frac{(N-2)(2N+1)}{2 N^{3/2} (N-1)} \bP_{a_1 a_2 a_3 , b_1 b_2 b_3} =: \lambda f_T (N) \bP_{a_1 a_2 a_3 , b_1 b_2 b_3} \,.
\eeq
The important fact is that $f_T (N)$ is a series in $\frac{1}{\sqrt{N}}$ with no constant term. 

\subsection{Elementary melon maps}

We now turn to elementary melon maps. There are six distinct two-point melon maps on two vertices:
\begin{center}
\includegraphics[scale=.7]{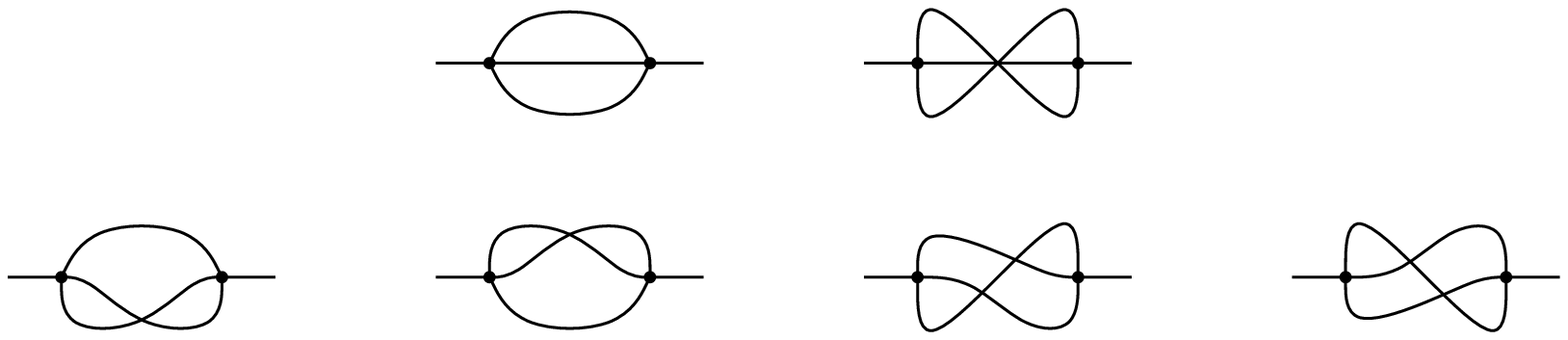}
\end{center} 
Due to the invariance of the vertex under cyclic reversal, the melon maps in a same row have the same amplitude. 

The amplitude of a melon from the first row is
\beq
A(\vcenter{\hbox{\includegraphics[scale=0.4]{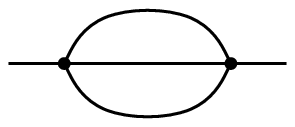}}}) = \lambda^2 \frac{8 N^6 - 24 N^5 - 23 N^4 + 58 N^3 + 120 N^2 - 88 N - 132}{216 \left( N- 1 \right)^3 N^3} \bP\,,
\eeq
while a melon from the second row evaluates to
\beq
A(\vcenter{\hbox{\includegraphics[scale=0.4]{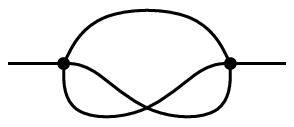}}}) =
\lambda^2 \frac{N^6 - 3 N^5 + 5 N^4 - 58 N^3 + 105 N^2 + 277 N - 246}{432 N^3 \left( N-1\right)^3} \bP\,.
\eeq
In contrast to tadpoles, melons have contributions of order $1$, they will therefore contribute to the leading-order in the large $N$ regime. Their sum is:
\beq
2 A(\vcenter{\hbox{\includegraphics[scale=0.4]{planar-melon}}}) + 4 A(\vcenter{\hbox{\includegraphics[scale=0.4]{non-planar-melon}}}) = \lambda^2 \frac{N^5 - 2 N^4 - 4 N^3 - 4 N^2 + 21 N + 42}{12 N^3 \left( N-1\right)^2}  \bP := \lambda^2 f_M (N) \bP\,,
\eeq
where $f_M (N)$ is a series in $1/N$ with non-vanishing constant term. 

\subsection{Melon-tadpole two-point function}

\begin{figure}[htb]
 \begin{center}
\includegraphics[scale=.7]{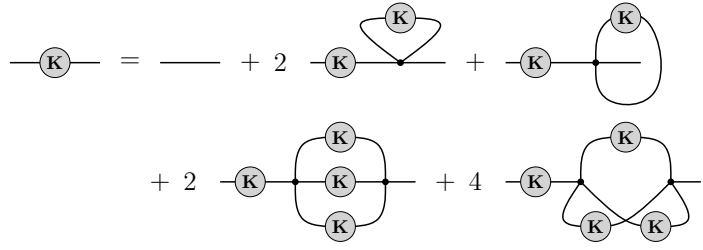}
 \caption{The Schwinger-Dyson equation of the melon-tadpole two-point function $\mathbf{K}$.} \label{fig:SDE}
 \end{center}
 \end{figure}

As a direct consequence of the recursive definition of melon-tadpoles, the two-point function $\mathbf{K}$ verifies the Schwinger--Dyson equation of Fig.~\ref{fig:SDE}. 

Furthermore, we already know that $\mathbf{K} = K \bP$ for some function $K(\lambda, N)$. As a result, the Schwinger--Dyson equation reduces to the simple algebraic equation:
\beq\label{eq:SDE}
K = 1 + \lambda  f_T (N) K^2 + \lambda^2 f_M (N) K^4\,.
\eeq 
As $f_T (N) \underset{N \to + \infty}{\rightarrow} 0$ and $f_M (N) \underset{N \to + \infty}{\rightarrow} 1/12$, this equation always has a non-perturbative solution (with $K(\lambda, N) \to 1$ when $\lambda \to 0$) in the large $N$ regime, and for sufficiently small $\vert \lambda \vert$. $K(\lambda, N)$ can furthermore be expanded as a series in $\frac{1}{\sqrt{N}}$. Keeping the leading term only, one obtains
\beq
K(\lambda , N) = K_0 (\lambda) + \mathrm{O}\left(\frac{1}{\sqrt{N}}\right)\,,
\eeq
where $K_0 (\lambda)$ is the solution of:
\beq
K_0 = 1 + \frac{\lambda^2}{12} {K_0}^4 \,.
\eeq
Remarkably, $K_0 (\lambda)$ does not receive any contribution from $f_T$; it reduces to  a sum over melonic two-point amplitudes. 

\section{Large N expansion}\label{sec:largeN}

\subsection{Existence of the large N expansion}

Having managed to resum the melon-tadpole family non-perturbatively, we can now rely on an improved perturbative expansion, around the free theory associated to the Gaussian measure $\extd\mu_\mathbf{K}$. 

For the connected two-point function and its trace, this gives:
\begin{align}
C_{a_1 a_2 a_3 , b_1 b_2 b_3} &=  \sum_{\substack{ \mathrm{connected} \;  2\mathrm{-point} \; \mathrm{maps} \;\cM \\ \mathrm{without}\; \mathrm{melon} \\ \mathrm{without} \; \mathrm{tadpole}}} \lambda^{V(\cM)} \,  K(\lambda,N)^{E(\cM)} \, A(\cM)_{a_1 a_2 a_3 , b_1 b_2 b_3} \label{eq:improved}  \\
C - K(\lambda , N) \frac{N \left( N^2 - 4 \right)}{3} &=  \sum_{\substack{\mathrm{connected}\; \mathrm{rooted} \; \mathrm{maps}\;\cM \\ \mathrm{without}\; \mathrm{melon} \\ \mathrm{without} \; \mathrm{tadpole}}} \lambda^{V(\cM)} \,  K(\lambda,N)^{2 V(\cM)} \, A(\cM) \label{eq:improved2}
\end{align}

There are only two differences with respect to \eqref{connected2point} and \eqref{trace2point}: 1) the propagator is $\mathbf{K} = K \bP$; 2) the Feynman maps being summed over have no melons and no tadpoles. The beauty of this reformulation is that it makes the existence of the large $N$ expansion explicit. 

Indeed, since we know from Proposition \ref{propo1} that $\omega(G) \geq 0$ for any melon-tadpole free stranded graph, we are guaranteed that $A(\cM)$ expands as $N^3$ times a (formal) series in $1/\sqrt{N}$, for any melon-tadpole-free vacuum map $\cM$. Together with the fact that $K(\lambda , N)$ is itself a series in $1/\sqrt{N}$, this allows to conclude that both $C$ and $C_{a_1 a_2 a_3 , b_1 b_2 b_3}$ have a well-defined large $N$ expansion. The leading order of $C$ is of order $N^3$, while that of $C_{a_1 a_2 a_3 , b_1 b_2 b_3}$ is of order $N^0$, and both expansions are governed by the parameter $\frac{1}{\sqrt{N}}$.

\subsection{Dominance of melon diagrams}

We now prove that no Feynman map other than melons contribute to the leading order two-point function. A natural conjecture would be that $\omega( G ) \geq 1/2$ for any non-melonic stranded diagram $G$. However, how was realized in \cite{Benedetti:2017qxl}, this conjecture does not survive closer inspection: there exists maps $\cM$ which are not melonic, but have vanishing degree stranded configurations. A crucial feature of such maps is that they all contain a generalized tadpole. This tells us that the same type of non-trivial cancellations identified in tadpole maps will also be at play here. 

\begin{figure}[htb]
 \begin{center}
\includegraphics[scale=.7]{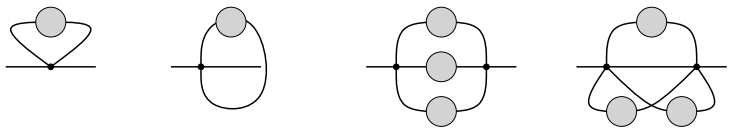}
 \caption{Generalized tadpoles and generalized melons.} \label{fig:gen-tadmel}
 \end{center}
 \end{figure}

In more detail, it is convenient to organize the discussion according to the presence or absence of generalized melons and generalized tadpoles (see Fig. \ref{fig:gen-tadmel}). Indeed, knowing that the large $N$ expansion of \eqref{eq:improved} exists, we can be sure that the large $N$ scaling of a generalized melon (resp. a generalized tadpole) is no higher than that of a melon (resp. a tadpole). Furthermore, the in-depth combinatorial investigation of \cite{Benedetti:2017qxl} leads to the result of Proposition \ref{propo1} \emph{ii)}, which we will rely on.  
  
\

Consider a two-point map $\cM$ in the improved perturbative expansion, that is: $\cM$ is connected, without melon and without tadpole. Let us furthermore suppose that $V(\cM) \geq 1$. We want to prove that $\cM$ cannot be leading-order.

We first argue that if $\cM$ has a generalized tadpole, then it cannot be leading order. Call $\cT$ the two-point map attached to the tadpole (which we assume to be planar for definiteness). Then $A(\cT) = a(\lambda,N) \bP$ with $a(\lambda , N)$ at most of order $N^0$ (otherwise the $1/N$ expansion would not exist: one could construct a graph with arbitrarily high scaling in $N$ by chaining $\cT$). But then $A( \vcenter{\hbox{\includegraphics[scale=0.4]{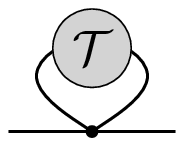}}}   ) \sim a(\lambda, N)  \frac{\lambda}{12 \sqrt{N}} \bP$. Calling $\cM'$ the map obtained from $\cM$ by substituting the generalized tadpole $\vcenter{\hbox{\includegraphics[scale=0.4]{tadpole-T}}}$ with a simple propagator, one has $A( \cM   ) \sim a(\lambda, N)  \frac{\lambda}{12 \sqrt{N}} A( \cM' ) $. Since $A(\cM' )$ scales at most as $N^0$, one concludes that $A(\cM)$ must decay as $1/\sqrt{N}$ or faster. 

Assume now that $\cM$ is leading order. By the previous argument together with Proposition \ref{propo1} \emph{ii)}, it follows that  $\cM$ must contain a generalized melon. Let $\tilde{\cM}$ be a generalized melon submap of $\cM$ with a minimal number of vertices. Up to orientation reversal of the vertices, $\tilde{M}$ has the structure:
\begin{center}
\includegraphics[scale=.7]{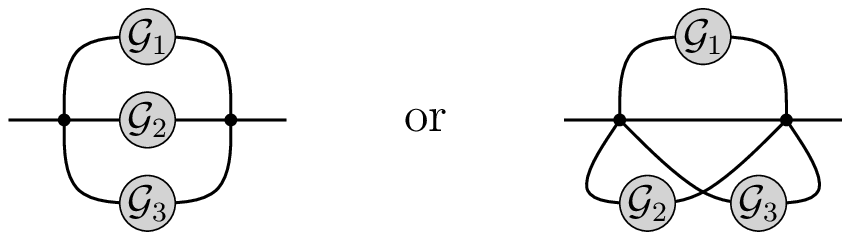}
\end{center}
with at least one of the subgraphs $\cG_i$ non-empty (otherwise $\cM$ would have a melon) -- say $\cG_1$. In order to ensure that $\cM$ is leading order, it is easy to see that $\cG_1$ must itself be leading order. But then it must also contain a generalized melon, which leads to a contradiction given that $\tilde{\cM}$ is a generalized melon with minimal number of vertices. Hence $\cM$ cannot be leading order, as we were claiming. 

We conclude that all the Feynman maps in the expansion \eqref{eq:improved} are suppressed by powers of $1/\sqrt{N}$, except for the dressed propagator $\mathbf{K} = K \bP$. Since we have shown that $K$ is itself a series in $1/\sqrt{N}$ whose constant term $K_0$ is a sum of melonic maps, we may conclude that the full connected two-point function converges to the melonic two-point function in the large $N$ limit. 

\

In equation, we have shown that:
\beq\label{eq:final}
\langle T_{a_1 a_2 a_3} T_{b_1 b_2 b_3} \rangle_{\mathrm{c}} = \left( K_0 (\lambda)  + \mathrm{O}( 1/ \sqrt{N} ) \right) \bP_{a_1 a_2 a_3, b_1 b_2 b_3} \,.
\eeq
where $K_0 (\lambda)$ is the sum of melonic two-point maps. In the $0$-dimensional context of the present paper, $K_0$ is furthermore a solution of a simple algebraic equation, namely $K_0 = 1 + \lambda^2 {K_0}^4 / 12$ (it is even exactly solvable, see for instance \cite{critical}).

\section{Conclusion}

We have proven that irreducible $O(N)$ tensor models associated to Young tableaux with the shape ${\vcenter{\hbox{\begin{Young}    &   \cr  \cr \end{Young}}}}$ admit a melonic large $N$ limit. Together with the results of \cite{Benedetti:2017qxl}, this achieves the proof that all irreducible rank-$3$ $O(N)$ tensor models with complete interaction can be organized into a large $N$ expansion dominated by melons. More generally, we conjecture that our construction can be extended to higher rank irreducible tensors: in rank $r$, one could consider a complete $(r+1)$-valent interaction. We hope to come back to this question in the near future. 

It would also be interesting to investigate the vector-tensor models which naturally follow from the decomposition of general tensors we relied on in section \ref{sec:vectormodes}. It was invoked to emphasize the instability introduced by the trace modes of the tensor, which we cured by removing these trace modes altogether. An alternative strategy would consist in introducing new independent 't Hooft couplings parametrizing the various types of interactions generated by the complete tensor interaction. In this way, one might presumably be able to construct a large $N$ expansion with a leading order including mixtures of melonic and bubble diagrams. 

Finally, we would like to emphasize that, even though we restricted our attention to a $0$-dimensional model, the methods of \cite{Benedetti:2017qxl} and of the present paper can be applied more generally to bosonic tensor field theories in $d \geq 1$. Provided that the global $O(N)$ symmetry is not spontaneously broken, an equation analogous to \eqref{eq:final} will automatically follow. The main difference is that the Schwinger--Dyson equation of the melonic two-point function determining $K_0$ will not reduce to a simple algebraic equation; it will result in a more involved melonic integro-differential equation, similar to those found in \cite{Klebanov:2016xxf, Giombi:2017dtl, Benedetti:2017fmp} (in $d=1$ and higher).

\section*{Acknowledgements}

\noindent I would like to thank Dario Benedetti, Razvan Gurau and Igor Klebanov for useful discussions.  

\noindent This research was supported in part by Perimeter Institute for Theoretical Physics. Research at Perimeter Institute is supported by the Government of Canada 
through the Department of Innovation, Science and Economic Development Canada and by the Province of Ontario through the Ministry of Research, Innovation and Science.

\let\oldbibliography\thebibliography
\renewcommand{\thebibliography}[1]{\oldbibliography{#1}
\setlength{\itemsep}{-1pt}}
\bibliographystyle{JHEP}
\bibliography{Refs.bib}

\providecommand{\href}[2]{#2}\begingroup\raggedright\begin{thebibliography}{10}

\bibitem{Benedetti:2017qxl}
D.~Benedetti, S.~Carrozza, R.~Gurau and M.~Kolanowski, \emph{{The $1/N$
  expansion of the symmetric traceless and the antisymmetric tensor models in
  rank three}},  \href{https://arxiv.org/abs/1712.00249}{{\tt 1712.00249}}.

\bibitem{expansion1}
R.~Gurau, \emph{{The {$1/N$} expansion of colored tensor models}},
  \href{http://dx.doi.org/10.1007/s00023-011-0101-8}{\emph{Ann. H. Poincar\'e}
  {\bf 12} (2011) 829--847}, [\href{https://arxiv.org/abs/1011.2726}{{\tt
  1011.2726}}].

\bibitem{expansion2}
R.~Gurau and V.~Rivasseau, \emph{{The {$1/N$} expansion of colored tensor
  models in arbitrary dimension}},
  \href{http://dx.doi.org/10.1209/0295-5075/95/50004}{\emph{Europhys. Lett.}
  {\bf 95} (2011) 50004}, [\href{https://arxiv.org/abs/1101.4182}{{\tt
  1101.4182}}].

\bibitem{expansion3}
R.~Gurau, \emph{{The complete {$1/N$} expansion of colored tensor models in
  arbitrary dimension}},
  \href{http://dx.doi.org/10.1007/s00023-011-0118-z}{\emph{Ann. H. Poincar\'e}
  {\bf 13} (2012) 399--423}, [\href{https://arxiv.org/abs/1102.5759}{{\tt
  1102.5759}}].

\bibitem{uncoloring}
V.~Bonzom, R.~Gurau and V.~Rivasseau, \emph{{Random tensor models in the large
  {$N$} limit: Uncoloring the colored tensor models}},
  \href{http://dx.doi.org/10.1103/PhysRevD.85.084037}{\emph{Phys. Rev.} {\bf
  D85} (2012) 084037}, [\href{https://arxiv.org/abs/1202.3637}{{\tt
  1202.3637}}].

\bibitem{expansion4}
V.~Bonzom, \emph{{New {$1/N$} expansions in random tensor models}},
  \href{http://dx.doi.org/10.1007/JHEP06(2013)062}{\emph{JHEP} {\bf 1306}
  (2013) 062}, [\href{https://arxiv.org/abs/1211.1657}{{\tt 1211.1657}}].

\bibitem{expansioin5}
S.~Dartois, V.~Rivasseau and A.~Tanasa, \emph{{The {$1/N$} expansion of
  multi-orientable random tensor models}},
  \href{http://dx.doi.org/10.1007/s00023-013-0262-8}{\emph{Ann. H. Poincar\'e}
  {\bf 15} (2014) 965--984}, [\href{https://arxiv.org/abs/1301.1535}{{\tt
  1301.1535}}].

\bibitem{expansioin6}
R.~Gurau, \emph{{The {$1/N$} expansion of tensor models beyond perturbation
  theory}}, \href{http://dx.doi.org/10.1007/s00220-014-1907-2}{\emph{Commun.
  Math. Phys.} {\bf 330} (2014) 973--1019},
  [\href{https://arxiv.org/abs/1304.2666}{{\tt 1304.2666}}].

\bibitem{Carrozza:2015adg}
S.~Carrozza and A.~Tanasa, \emph{{$O(N)$ Random Tensor Models}},
  \href{http://dx.doi.org/10.1007/s11005-016-0879-x}{\emph{Lett. Math. Phys.}
  {\bf 106} (2016) 1531--1559}, [\href{https://arxiv.org/abs/1512.06718}{{\tt
  1512.06718}}].

\bibitem{critical}
V.~Bonzom, R.~Gurau, A.~Riello and V.~Rivasseau, \emph{{Critical behavior of
  colored tensor models in the large {$N$} limit}},
  \href{http://dx.doi.org/10.1016/j.nuclphysb.2011.07.022}{\emph{Nucl. Phys.}
  {\bf B853} (2011) 174--195}, [\href{https://arxiv.org/abs/1105.3122}{{\tt
  1105.3122}}].

\bibitem{review}
R.~Gurau and J.~P. Ryan, \emph{{Colored tensor models - a review}},
  \href{http://dx.doi.org/10.3842/SIGMA.2012.020}{\emph{SIGMA} {\bf 8} (2012)
  020}, [\href{https://arxiv.org/abs/1109.4812}{{\tt 1109.4812}}].

\bibitem{RTM}
R.~Gurau, \emph{{Random Tensors}}.
\newblock Oxford University Press, Oxford, 2016.

\bibitem{EDT}
D.~Benedetti and R.~Gurau, \emph{{Phase transition in dually weighted colored
  tensor models}},
  \href{http://dx.doi.org/10.1016/j.nuclphysb.2011.10.015}{\emph{Nucl. Phys.}
  {\bf B855} (2012) 420--437}, [\href{https://arxiv.org/abs/1108.5389}{{\tt
  1108.5389}}].

\bibitem{IsingD}
V.~Bonzom, R.~Gurau and V.~Rivasseau, \emph{{The {Ising} model on random
  lattices in arbitrary dimensions}},
  \href{http://dx.doi.org/10.1016/j.physletb.2012.03.054}{\emph{Phys. Lett.}
  {\bf B711} (2012) 88--96}, [\href{https://arxiv.org/abs/1108.6269}{{\tt
  1108.6269}}].

\bibitem{melbp}
R.~Gurau and J.~P. Ryan, \emph{{Melons are branched polymers}},
  \href{http://dx.doi.org/10.1007/s00023-013-0291-3}{\emph{Ann. H. Poincar\'e}
  {\bf 15} (2014) 2085--2131}, [\href{https://arxiv.org/abs/1302.4386}{{\tt
  1302.4386}}].

\bibitem{Benedetti:2015ara}
D.~Benedetti and R.~Gurau, \emph{{Symmetry breaking in tensor models}},
  \href{http://dx.doi.org/10.1103/PhysRevD.92.104041}{\emph{Phys. Rev.} {\bf
  D92} (2015) 104041}, [\href{https://arxiv.org/abs/1506.08542}{{\tt
  1506.08542}}].

\bibitem{GurSch}
R.~Gurau and G.~Schaeffer, \emph{{Regular colored graphs of positive degree}},
  \href{http://dx.doi.org/10.4171/AIHPD/29}{\emph{Ann. Inst. Henri Poincar\'e
  Comb. Phys. Interact.} {\bf 3} (2016) 257--320},
  [\href{https://arxiv.org/abs/1307.5279}{{\tt 1307.5279}}].

\bibitem{BenGeloun:2011rc}
J.~Ben~Geloun and V.~Rivasseau, \emph{{A renormalizable 4-Dimensional tensor
  field theory}},
  \href{http://dx.doi.org/10.1007/s00220-012-1549-1}{\emph{Commun. Math. Phys.}
  {\bf 318} (2013) 69--109}, [\href{https://arxiv.org/abs/1111.4997}{{\tt
  1111.4997}}].

\bibitem{Carrozza:2012uv}
S.~Carrozza, D.~Oriti and V.~Rivasseau, \emph{{Renormalization of Tensorial
  Group Field Theories: {Abelian} {$U(1)$} Models in Four Dimensions}},
  \href{http://dx.doi.org/10.1007/s00220-014-1954-8}{\emph{Commun. Math. Phys.}
  {\bf 327} (2014) 603--641}, [\href{https://arxiv.org/abs/1207.6734}{{\tt
  1207.6734}}].

\bibitem{Samary:2012bw}
D.~O. Samary and F.~Vignes-Tourneret, \emph{{Just Renormalizable {TGFT}'s on
  {$U(1)^{d}$} with Gauge Invariance}},
  \href{http://dx.doi.org/10.1007/s00220-014-1930-3}{\emph{Commun. Math. Phys.}
  {\bf 329} (2014) 545--578}, [\href{https://arxiv.org/abs/1211.2618}{{\tt
  1211.2618}}].

\bibitem{tt2}
S.~Carrozza, D.~Oriti and V.~Rivasseau, \emph{{Renormalization of a {SU(2)}
  Tensorial Group Field Theory in Three Dimensions}},
  \href{http://dx.doi.org/10.1007/s00220-014-1928-x}{\emph{Commun. Math. Phys.}
  {\bf 330} (2014) 581--637}, [\href{https://arxiv.org/abs/1303.6772}{{\tt
  1303.6772}}].

\bibitem{Baratin:2013rja}
A.~Baratin, S.~Carrozza, D.~Oriti, J.~P. Ryan and M.~Smerlak, \emph{{Melonic
  phase transition in group field theory}},
  \href{http://dx.doi.org/10.1007/s11005-014-0699-9}{\emph{Lett. Math. Phys.}
  {\bf 104} (2014) 1003--1017}, [\href{https://arxiv.org/abs/1307.5026}{{\tt
  1307.5026}}].

\bibitem{Carrozza:2014rba}
S.~Carrozza, \emph{{Discrete Renormalization Group for {SU(2)} Tensorial Group
  Field Theory}}, \href{http://dx.doi.org/10.4171/AIHPD/15}{\emph{Ann. Inst.
  Henri Poincar\'e Comb. Phys. Interact.} {\bf 2} (2015) 49--112},
  [\href{https://arxiv.org/abs/1407.4615}{{\tt 1407.4615}}].

\bibitem{Delepouve:2015nia}
T.~Delepouve and R.~Gurau, \emph{{Phase Transition in Tensor Models}},
  \href{http://dx.doi.org/10.1007/JHEP06(2015)178}{\emph{JHEP} {\bf 06} (2015)
  178}, [\href{https://arxiv.org/abs/1504.05745}{{\tt 1504.05745}}].

\bibitem{Benedetti:2015yaa}
D.~Benedetti and V.~Lahoche, \emph{{Functional Renormalization Group Approach
  for Tensorial Group Field Theory: A Rank-6 Model with Closure Constraint}},
  \href{http://dx.doi.org/10.1088/0264-9381/33/9/095003}{\emph{Class. Quant.
  Grav.} {\bf 33} (2016) 095003}, [\href{https://arxiv.org/abs/1508.06384}{{\tt
  1508.06384}}].

\bibitem{Witten:2016iux}
E.~Witten, \emph{{An SYK-Like Model Without Disorder}},
  \href{https://arxiv.org/abs/1610.09758}{{\tt 1610.09758}}.

\bibitem{Sachdev:1992fk}
S.~Sachdev and J.~Ye, \emph{{Gapless spin fluid ground state in a random,
  quantum Heisenberg magnet}},
  \href{http://dx.doi.org/10.1103/PhysRevLett.70.3339}{\emph{Phys. Rev. Lett.}
  {\bf 70} (1993) 3339}, [\href{https://arxiv.org/abs/cond-mat/9212030}{{\tt
  cond-mat/9212030}}].

\bibitem{Kitaev}
A.~Kitaev, \emph{{A simple model of quantum holography}}, {\emph{KITP strings
  seminar and Entanglement 2015 program (Feb. 12, April 7, and May 27, 2015)}
  }.

\bibitem{Maldacena:2016hyu}
J.~Maldacena and D.~Stanford, \emph{{Remarks on the Sachdev-Ye-Kitaev model}},
  \href{http://dx.doi.org/10.1103/PhysRevD.94.106002}{\emph{Phys. Rev.} {\bf
  D94} (2016) 106002}, [\href{https://arxiv.org/abs/1604.07818}{{\tt
  1604.07818}}].

\bibitem{Polchinski:2016xgd}
J.~Polchinski and V.~Rosenhaus, \emph{{The Spectrum in the Sachdev-Ye-Kitaev
  Model}}, \href{http://dx.doi.org/10.1007/JHEP04(2016)001}{\emph{JHEP} {\bf
  04} (2016) 001}, [\href{https://arxiv.org/abs/1601.06768}{{\tt 1601.06768}}].

\bibitem{Klebanov:2016xxf}
I.~R. Klebanov and G.~Tarnopolsky, \emph{{Uncolored Random Tensors, Melon
  Diagrams, and the SYK Models}},
  \href{http://dx.doi.org/10.1103/PhysRevD.95.046004}{\emph{Phys. Rev.} {\bf
  D95} (2017) 046004}, [\href{https://arxiv.org/abs/1611.08915}{{\tt
  1611.08915}}].

\bibitem{Gurau:2016lzk}
R.~Gurau, \emph{{The complete $1/N$ expansion of a SYK--like tensor model}},
  \href{http://dx.doi.org/10.1016/j.nuclphysb.2017.01.015}{\emph{Nucl. Phys.}
  {\bf B916} (2017) 386--401}, [\href{https://arxiv.org/abs/1611.04032}{{\tt
  1611.04032}}].

\bibitem{Krishnan:2016bvg}
C.~Krishnan, S.~Sanyal and P.~N. Bala~Subramanian, \emph{{Quantum Chaos and
  Holographic Tensor Models}},
  \href{http://dx.doi.org/10.1007/JHEP03(2017)056}{\emph{JHEP} {\bf 03} (2017)
  056}, [\href{https://arxiv.org/abs/1612.06330}{{\tt 1612.06330}}].

\bibitem{Krishnan:2017ztz}
C.~Krishnan, K.~V.~P. Kumar and S.~Sanyal, \emph{{Random Matrices and
  Holographic Tensor Models}},
  \href{http://dx.doi.org/10.1007/JHEP06(2017)036}{\emph{JHEP} {\bf 06} (2017)
  036}, [\href{https://arxiv.org/abs/1703.08155}{{\tt 1703.08155}}].

\bibitem{Krishnan:2017lra}
C.~Krishnan, K.~V. Pavan~Kumar and D.~Rosa, \emph{{Contrasting SYK-like
  Models}}, \href{http://dx.doi.org/10.1007/JHEP01(2018)064}{\emph{JHEP} {\bf
  01} (2018) 064}, [\href{https://arxiv.org/abs/1709.06498}{{\tt 1709.06498}}].

\bibitem{Bonzom:2017pqs}
V.~Bonzom, L.~Lionni and A.~Tanasa, \emph{{Diagrammatics of a colored SYK model
  and of an SYK-like tensor model, leading and next-to-leading orders}},
  \href{http://dx.doi.org/10.1063/1.4983562}{\emph{J. Math. Phys.} {\bf 58}
  (2017) 052301}, [\href{https://arxiv.org/abs/1702.06944}{{\tt 1702.06944}}].

\bibitem{Bulycheva:2017ilt}
K.~Bulycheva, I.~R. Klebanov, A.~Milekhin and G.~Tarnopolsky, \emph{{Spectra of
  Operators in Large $N$ Tensor Models}},
  \href{http://dx.doi.org/10.1103/PhysRevD.97.026016}{\emph{Phys. Rev.} {\bf
  D97} (2018) 026016}, [\href{https://arxiv.org/abs/1707.09347}{{\tt
  1707.09347}}].

\bibitem{Benedetti:2018goh}
D.~Benedetti and R.~Gurau, \emph{{2PI effective action for the SYK model and
  tensor field theories}},
  \href{http://dx.doi.org/10.1007/JHEP05(2018)156}{\emph{JHEP} {\bf 05} (2018)
  156}, [\href{https://arxiv.org/abs/1802.05500}{{\tt 1802.05500}}].

\bibitem{Klebanov:2018nfp}
I.~R. Klebanov, A.~Milekhin, F.~Popov and G.~Tarnopolsky, \emph{{On the Spectra
  of Eigenstates in Fermionic Tensor Quantum Mechanics}},
  \href{https://arxiv.org/abs/1802.10263}{{\tt 1802.10263}}.

\bibitem{Giombi:2017dtl}
S.~Giombi, I.~R. Klebanov and G.~Tarnopolsky, \emph{{Bosonic tensor models at
  large $N$ and small $\epsilon$}},
  \href{http://dx.doi.org/10.1103/PhysRevD.96.106014}{\emph{Phys. Rev.} {\bf
  D96} (2017) 106014}, [\href{https://arxiv.org/abs/1707.03866}{{\tt
  1707.03866}}].

\bibitem{Prakash:2017hwq}
S.~Prakash and R.~Sinha, \emph{{A Complex Fermionic Tensor Model in $d$
  Dimensions}}, \href{http://dx.doi.org/10.1007/JHEP02(2018)086}{\emph{JHEP}
  {\bf 02} (2018) 086}, [\href{https://arxiv.org/abs/1710.09357}{{\tt
  1710.09357}}].

\bibitem{Benedetti:2017fmp}
D.~Benedetti, S.~Carrozza, R.~Gurau and A.~Sfondrini, \emph{{Tensorial
  Gross-Neveu models}},
  \href{http://dx.doi.org/10.1007/JHEP01(2018)003}{\emph{JHEP} {\bf 01} (2018)
  003}, [\href{https://arxiv.org/abs/1710.10253}{{\tt 1710.10253}}].

\bibitem{BenGeloun:2017jbi}
J.~Ben~Geloun and V.~Rivasseau, \emph{{A Renormalizable SYK-type Tensor Field
  Theory}},  \href{https://arxiv.org/abs/1711.05967}{{\tt 1711.05967}}.

\bibitem{Choudhury:2017tax}
S.~Choudhury, A.~Dey, I.~Halder, L.~Janagal, S.~Minwalla and R.~Poojary,
  \emph{{Notes on Melonic $O(N)^{q-1}$ Tensor Models}},
  \href{https://arxiv.org/abs/1707.09352}{{\tt 1707.09352}}.

\bibitem{Ferrari:2017ryl}
F.~Ferrari, \emph{{The Large D Limit of Planar Diagrams}},
  \href{https://arxiv.org/abs/1701.01171}{{\tt 1701.01171}}.

\bibitem{Azeyanagi:2017drg}
T.~Azeyanagi, F.~Ferrari and F.~I. Schaposnik~Massolo, \emph{{Phase Diagram of
  Planar Matrix Quantum Mechanics, Tensor, and Sachdev-Ye-Kitaev Models}},
  \href{http://dx.doi.org/10.1103/PhysRevLett.120.061602}{\emph{Phys. Rev.
  Lett.} {\bf 120} (2018) 061602},
  [\href{https://arxiv.org/abs/1707.03431}{{\tt 1707.03431}}].

\bibitem{Ferrari:2017jgw}
F.~Ferrari, V.~Rivasseau and G.~Valette, \emph{{A New Large N Expansion for
  General Matrix-Tensor Models}},  \href{https://arxiv.org/abs/1709.07366}{{\tt
  1709.07366}}.

\bibitem{Azeyanagi:2017mre}
T.~Azeyanagi, F.~Ferrari, P.~Gregori, L.~Leduc and G.~Valette, \emph{{More on
  the New Large $D$ Limit of Matrix Models}},
  \href{https://arxiv.org/abs/1710.07263}{{\tt 1710.07263}}.

\bibitem{Ryan:2016sundry}
J.~P. Ryan, \emph{{{(D+1)-Colored Graphs - a Review of Sundry Properties}}},
  {\emph{SIGMA} {\bf 12} (2016) 076}.

\bibitem{Klebanov:2017nlk}
I.~R. Klebanov and G.~Tarnopolsky, \emph{{On Large $N$ Limit of Symmetric
  Traceless Tensor Models}},
  \href{http://dx.doi.org/10.1007/JHEP10(2017)037}{\emph{JHEP} {\bf 10} (2017)
  037}, [\href{https://arxiv.org/abs/1706.00839}{{\tt 1706.00839}}].

\bibitem{Gurau:2017qya}
R.~Gurau, \emph{{The $1/N$ expansion of tensor models with two symmetric
  tensors}}, \href{http://dx.doi.org/10.1007/s00220-017-3055-y}{\emph{Commun.
  Math. Phys.} (2017) }, [\href{https://arxiv.org/abs/1706.05328}{{\tt
  1706.05328}}].

\bibitem{Moshe:2003xn}
M.~Moshe and J.~Zinn-Justin, \emph{{Quantum field theory in the large N limit:
  A Review}},
  \href{http://dx.doi.org/10.1016/S0370-1573(03)00263-1}{\emph{Phys. Rept.}
  {\bf 385} (2003) 69--228}, [\href{https://arxiv.org/abs/hep-th/0306133}{{\tt
  hep-th/0306133}}].

\bibitem{weyl1946}
H.~Weyl, \emph{The Classical Groups, Their Invariants and Representations}.
\newblock Princeton University Press, 1946.

\bibitem{hamermesh}
M.~Hamermesh, \emph{Group theory and its application to physical problems}.
\newblock Dover, 1989.

\bibitem{lando-zvonkin-graphs}
S.~K. Lando and A.~K. Zvonkin, \emph{Graphs on surfaces and their
  applications}, vol.~141 of \emph{Encyclopaedia of Mathematical Sciences}.
\newblock Springer, 2004.

\bibitem{ellis2013graphs}
J.~A. Ellis-Monaghan and I.~Moffatt, \emph{Graphs on surfaces: dualities,
  polynomials, and knots}, vol.~84.
\newblock Springer, 2013.

\end{thebibliography}\endgroup

\end{document}